\documentclass[aps,prb,english,twocolumn,showpacs,preprintnumbers,amsmath,amssymb,floatfix,superscriptaddress,longbibliography]{revtex4-1}

\usepackage[T1]{fontenc}
\usepackage[latin9]{inputenc}
\usepackage{graphicx}
\usepackage{amssymb}
\usepackage{babel}
\makeatletter

\usepackage[version=3]{mhchem}

\usepackage{color}

\usepackage{hyperref}
\hypersetup{
     colorlinks = true,
     citecolor  = blue,  
     linkcolor  = magenta 
}

\makeatother
\usepackage{babel}
\makeatother

\begin{document}

\title{Casimir-Lifshitz force variations across heterogeneous gapped metal surfaces}

\author{M.  Bostr{\"o}m}
  \email{mathias.bostrom@ensemble3.eu}
 \affiliation{Centre of Excellence ENSEMBLE3 Sp. z o. o., Wolczynska Str. 133, 01-919, Warsaw, Poland}
 \affiliation{Chemical and Biological Systems Simulation Lab, Centre of New Technologies, University of Warsaw, Banacha 2C, 02-097 Warsaw, Poland}

  \author{S. Pal}
  \affiliation{Centre of Excellence ENSEMBLE3 Sp. z o. o., Wolczynska Str. 133, 01-919, Warsaw, Poland}

 \author{H. R. Gopidi}
\affiliation{Centre of Excellence ENSEMBLE3 Sp. z o. o., Wolczynska Str. 133, 01-919, Warsaw, Poland}

  \author{S. Osella}
 \affiliation{Chemical and Biological Systems Simulation Lab, Centre of New Technologies, University of Warsaw, Banacha 2C, 02-097 Warsaw, Poland}

\author{A. Gholamhosseinian}
  \affiliation{Department of Physics, Ferdowsi University of Mashhad, Mashad, Iran}

\author{G. Palasantzas}
\affiliation{Zernike Institute for Advanced Materials, University of Groningen, Groningen, 9747 AG, The Netherlands}

\author{O. I. Malyi}
\email{oleksandrmalyi@gmail.com}
\affiliation{Centre of Excellence ENSEMBLE3 Sp. z o. o., Wolczynska Str. 133, 01-919, Warsaw, Poland}

\date{\today}%

\begin{abstract}
The Casimir-Lifshitz force is calculated between a heterogeneous gapped metal surface and a silica (gold) sphere attached to an AFM cantilever tip. We demonstrate that heterogeneous surface patches with different off-stoichiometry surface properties lead to changes in the predicted distances for a specific force. This can incorrectly be interpreted as occurrences of surface roughness.
\end{abstract}

\maketitle

\section{Introduction}

With the development of solid-state physics and quantum
theory, it becomes clear that the interaction between
closely spaced objects originate from the fluctuation of
the electromagnetic field defining the Casimir interaction\,\cite{Casi}. 
The fundamental physics of this interaction has
been developed for a range of simple geometries and is
defined by the dielectric properties of interacting
materials\,\cite{Dzya}. This foundation formed the basis
for our understanding of Casimir's effect for real metal
surfaces\,\cite{Ser2018,MostKlim2023,BostromRizwanHarshanBrevikLiPerssonMalyi2023spontaneous}, repulsive Casimir forces with topological insulators\,\cite{PhysRevLett.106.020403}, Casimir torque\,\cite{SomersGarrettPalmMunday_CasimirTorque},  and for cylinders across magnetic fluids\,\cite{NylandBrevik1994}. 
Force measurements have, for instance, been carried out using a torsion pendulum\,\cite{Lamo1997,SushNP} and with the help of atomic force microscope (AFM)\,\cite{DuckerSendenPashley_AFM_Nature}, e.g. in sphere-plate setups.
It is widely
believed that the dielectric properties of many materials
are weakly affected by the environmental conditions, since,
 under standard conditions,
the environmental effect is predominantly limited to the
concentration of defects within a specific material\,\cite{ZungerMalyiChemRev2021}.  
However, with the emergence of electronic structure
theory, a new type of materials - gapped metals -
has been identified. What makes these materials special
is that they exhibit the superposition of both insulating
and metallic properties, i.e., having an internal gap and
larger free carrier concentration due to the Fermi level residing in the principal conduction (i.e., n-type) or valence band (i.e., p-type). Such a unique electronic structure not only results in unique dielectric properties but also offers a knob that can be used to tune materials properties\,\cite{Malyi2019_doi:10.1016/j.matt.2019.05.014}.
Specifically, gapped metals can develop spontaneous
off-stoichiometry due to Fermi-level instability\,\cite{MalyiZUngerApplPhysRev2020}, resulting in the formation
of a range of off-stoichiometric compounds, all having different
electronic properties. Taking into account that the
dielectric properties of any given material are closely related
to electronic structure, it becomes clear that such
a knob can also be used to tune the Casimir interaction.

Moving from this theoretical framework to practical
investigation, atomic force microscopy (AFM) the technique of choice when
the Casimir interaction
play a dominant role. Consider, for instance, the case
of AFM measurements of materials roughness, where the
typical materials profile is extracted based on a specific
physical guess of involved interactions. 
In the case of homogeneous samples, such material profiles can be well
theoretically motivated. However, for commonly observed heterogeneous samples,
different regions of the
samples can result in different interactions with the tip.
The observed force is highly sensitive to both surface roughness and optical properties at separations less than 100\,nm, as reported by Broer{\it et al.}\,\cite{BroerPalasantzasPhysRevB.85.155410_2012}.
Motivated by this, in this letter, we explore
  the fundamental theory of the
Casimir interaction for gapped metal surfaces. Notably, our results clearly demonstrate how a change in stoichiometry can result in changes in the Casimir forces,
which can be misinterpreted as a change of roughness when not properly accounting for heterogeneous surface-specific properties.

\section{Theory}

To study and model the dielectric properties of the relevant materials, we performed first-principles calculations using the Perdew-Burke-Ernzerhof (PBE) exchange-correlation functional\,\cite{perdew1996generalized} with DFT+U correction for Nb (U = 1.5 eV) d-like orbitals as implemented by Dudarev {\it et al.}\,\cite{Durdev1998_doi:10.1103/PhysRevB.57.1505} available within the  Vienna Ab initio Simulation Package (VASP)\,\cite{vasp1,vasp3,vasp4,vasp1999}. Our analysis focuses on two representatives sets of gapped metals, (i) Ba$_{1-x}$Nb$_{1-y}$O$_{3}$ (with 5 different compositions) and (ii) Ca$_{6-x}$Al$_{7}$O$_{16}$ (with 3 different compositions) previously identified to be stable with respect to decomposition into the competing phases\,\cite{Malyi2019_doi:10.1016/j.matt.2019.05.014}. We compute the dielectric properties for each system, considering only the Drude contribution and interband transitions. For the calculations of direct band transitions and plasma frequencies we introduced a small Lorentzian broadening of 0.01 eV in the Kramers-Kronig transformation\,\cite{landau2013statistical}. To include the Drude term in the optical properties, we utilize the kram code (which is part of WIEN2k\,\cite{blaha2020wien2k, blaha1990full, blaha2001wien2k}) setting the damping coefficient ($\Gamma$) to 0.2 \,eV. Additional details on computational parameters can be found in our previous work\,\cite{Malyi2019_doi:10.1016/j.matt.2019.05.014}. 

\begin{figure*}
 \includegraphics[width=0.9\textwidth,height=6.8cm]{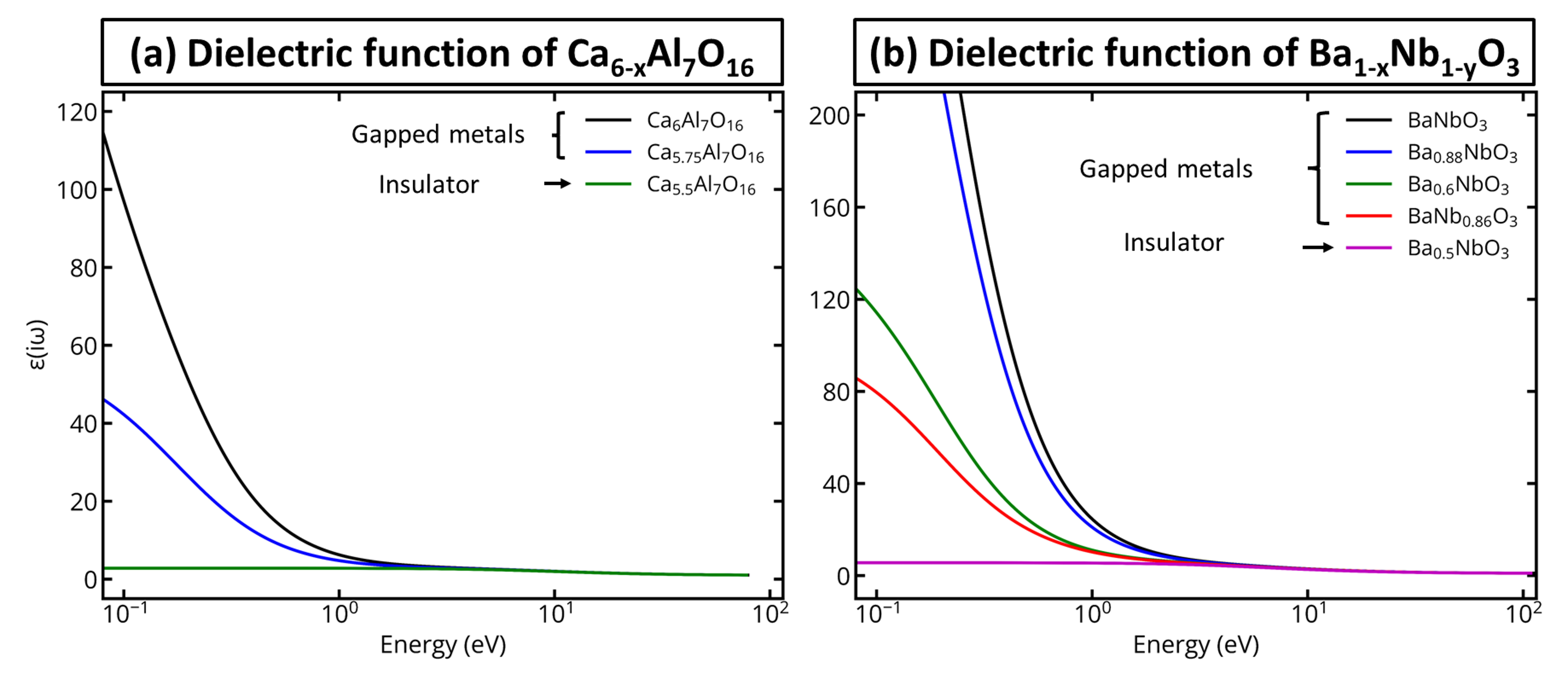}
 \vspace{-0.15in}
  \caption{\label{DielectricFunctions}   (Color online) (a) Dielectric functions for imaginary frequencies from top to bottom for Ca$_{6}$Al$_{7}$O$_{16}$, Ca$_{5.75}$Al$_{7}$O$_{16}$, and Ca$_{5.5}$Al$_{7}$O$_{16}$.  (b) Dielectric functions for imaginary frequencies from top to bottom for BaNbO$_{3}$, BaNb$_{0.88}$O$_{3}$, Ba$_{0.6}$NbO$_{3}$, BaNb$_{0.86}$O$_{3}$, and Ba$_{0.5}$NbO$_{3}$. The results include a superposition of interband transitions and Drude (free electron) contributions (with damping coefficient ($\Gamma$) set to 0.2 eV), both calculated using DFT.}
\end{figure*}

From the imaginary part ($\varepsilon_i''$ for material $i$) of the dielectric function, the quantity related to forces can be obtained:
\begin{equation}
\varepsilon_i(i \xi_m)=1+\frac{2}{\pi}\int_0^\infty d\omega \frac{ \omega \varepsilon_i''(\omega)}{\omega^2+\xi_m^2},\,{i=1,2,3}
\end{equation}
where the Matsubara frequency is $\xi_m=2 \pi k T m/\hbar$, and the subscript \textit{i} indicates the medium. As seen in Fig.\,\ref{DielectricFunctions} the curves show strong dependence for the dielectric function on off-stoichiometry for Ba$_{1-x}$Nb$_{1-y}$O$_{3}$ and Ca$_{6-x}$Al$_{7}$O$_{16}$ going from a metallic to insulator behavior. This behavior mainly originated from the spontaneous formation of cation vacancies, originating from energy lowering due to the decay of conducting electrons into the acceptor defect states suggesting that the tuning synthesis/environmental conditions allows to stabilize different gapped metals \cite{Malyi2019_doi:10.1016/j.matt.2019.05.014}.  To use the DFT calculated dielectric functions for Casimir-Lifshitz interaction, we also develop the parametrization of the average dielectric function for each of the considered compounds using a 13-mode oscillator model\,\cite{Malyi_etal_PCCP_VolDepDielAmorphousSiO2_2016}: 

\begin{equation}
\varepsilon(i \xi)=1+\sum_j \frac{C_{j}}{1+ (\xi/\omega_j)^2}.
        \label{ParameteriseddielEq}
\end{equation}
Here, $\omega_j$ and $C_j$ represent the characteristic frequencies and the oscillator strength, respectively. The parameters for each gapped metal are provided in Appendix.\,(\ref{appenb}).

Let $f(d)$ be the force between a gapped metal (medium 1) and a sphere with radius $R$ (medium 3).  For $d<<R$ the force between a sphere and a planar surface can be deduced from the free energy ($F(d)=f(d)/[2 \pi R]$) between two planar surfaces using the so-called proximity force approximation\,\cite{Ser2018}. The sphere is taken to be made of SiO$_2$ polymorph (with a volume per SiO$_2$ unit being 68.82 \AA$^3$ with the details in Ref. \cite{Malyi_etal_PCCP_VolDepDielAmorphousSiO2_2016} and including phonon contribution for low frequencies as done by  Bostr\"om {\it et al.}\,\cite{MBPhysRevB02017,LiMiltonBrevikMalyiThiyamPerssonParsonsBostrom_PRB2022}).  The intervening medium 2 may be a diluted gas with $\varepsilon_2(i \xi_m)=1$. The force can be written as\,\cite{Dzya,Ser2018},
\begin{equation}
\frac{f(d)}{2 \pi R} = \frac{k_BT}{2 \pi} {\sum_{m=0}^\infty}{}^\prime \int\limits_0^\infty dq\,q \sum_{\sigma}\ln(1- r_{\sigma}^{21}r_{\sigma}^{23}
 \mathrm e^{-2\kappa_2 d}), \label{LifFreeEnergy}
\end{equation}
where $\sigma=\rm TE,TM$, $k_B$ is Boltzmann's constant, $\hbar$ is Planck's constant, temperature is T=300\,K, and the prime in the sum above indicates that the first term ($m$ = 0) has to be weighted by $1/2$. The Fresnel reflection coefficients between surfaces $i$ and $j$ for the transverse magnetic (TM) and transverse electric (TE) polarizations are given by
\begin{equation}
    r_{\rm TE}^{ij} = \frac{\kappa_i-\kappa_j}{\kappa_i+\kappa_j}\,;  \,\,\,\,\, r_{\rm TM}^{ij} = \frac{\varepsilon_j\kappa_i-\varepsilon_i \kappa_j}{\varepsilon_j \kappa_i+\varepsilon_i \kappa_j} \,. \label{eq:rtTETM}
\end{equation}
Here $\kappa_i= \sqrt{{q}^2+\varepsilon_i\xi_m^2/c^2}$, with $i=1,2,3$ and the Matsubara frequency being $\xi_m=2 \pi k_B T m/\hbar$.

\section{Results}

\subsection{Silica systems}

\begin{figure}[h]
  \centering
  \includegraphics[width=1\columnwidth]{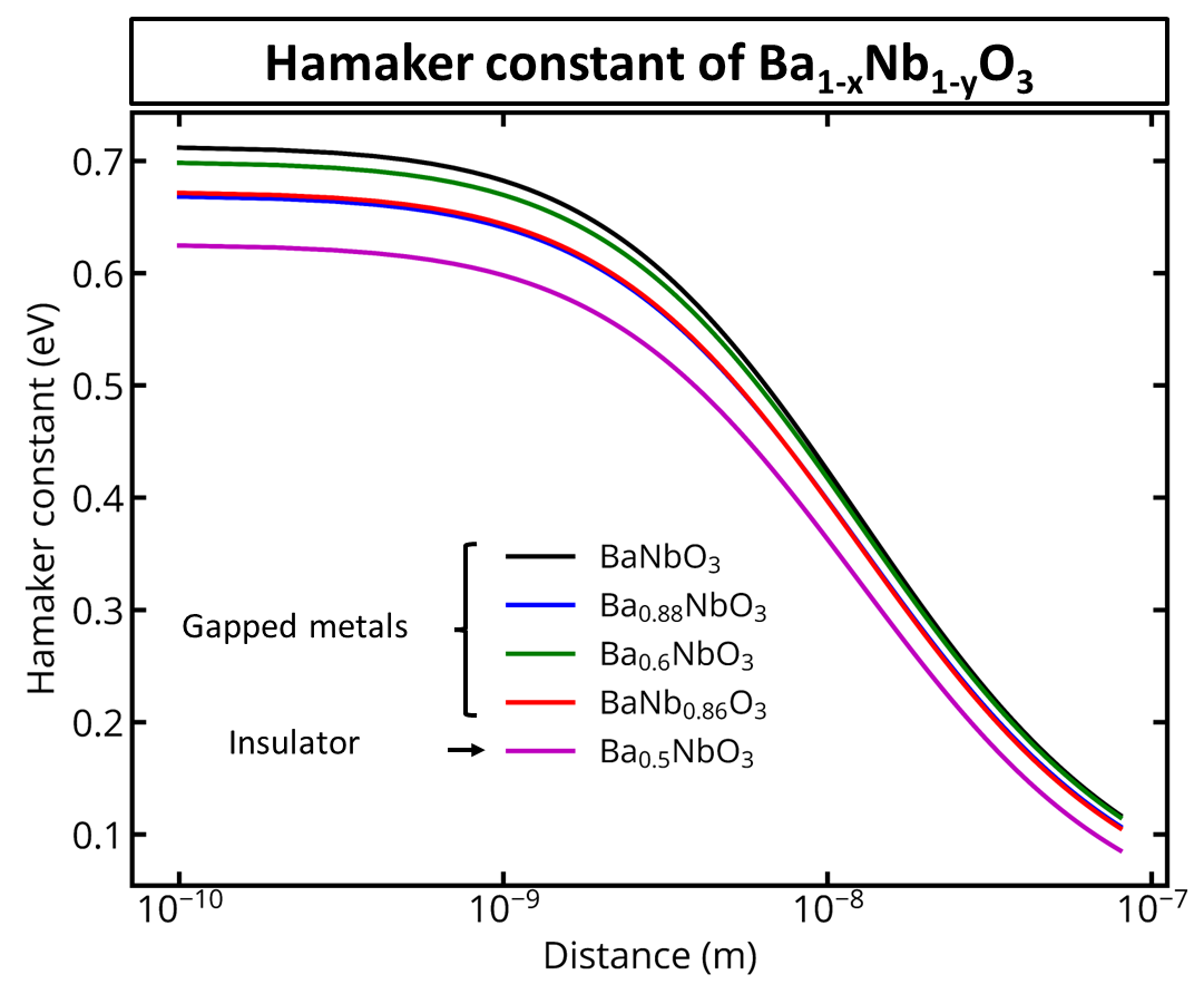}
   \vspace{-0.25in}
\caption{\label{RetardedHamaker1} (Color online) The retarded Hamaker constant $A^{\text{ret}}(d) = -12 \pi d^2 \times F(d, T)$ for several material combinations of Ba$_{1-x}$Nb$_{1-y}$O$_{3}-$air$-$SiO$_2$ at temperature $T = 300$ K.}
\end{figure}

\begin{table}
\centering
\begin{tabular}{c|c|c|c}
  \hline
   Material & $A^{NR}$ & $\delta_o^{NR}$ & $\delta_o$ at $d_s$=80\,nm\\
    \hline
    BaNbO$_{3}$ &  0.71 eV& x&x\\
    \hline
     Ba$_{0.88}$NbO$_{3}$  & 0.66  eV&  $ 3.71 \%$ &$ 0.56\%$\\
    \hline
     Ba$_{0.6}$NbO$_{3}$  &  0.69 eV & $ 1.43\%$ &$ 3.11\%$\\
    \hline
     BaNb$_{0.86}$O$_{3}$  &  0.67  eV& $  2.94 \%$ &$ 3.68\%$\\
    \hline
     Ba$_{0.5}$NbO$_{3}$  &  0.63 eV& $ 6.15 \%$ &$ 10.8\%$\\
    \hline
     \hline
      \hline
      Ca$_{6}$Al$_{7}$O$_{16}$ & 0.50  eV& x &x\\
    \hline
      Ca$_{5.75}$Al$_{7}$O$_{16}$ &   0.48 eV&$ 2.06 \%$ &$ 2.59\%$\\
    \hline
     Ca$_{5.5}$Al$_{7}$O$_{16}$  & 0.44 eV & $  6.6\%$ &$ 16.57\%$\\
    \hline
\end{tabular}
\caption{\label{tbFittingParametersSubhojit} The non-retarded Hamaker constant for gapped metal-vapor-SiO$_2$. With simple manipulation of equations the relative surface correction in the non-retarded limit can be shown equal to $\delta_o=100*(\sqrt{A^{NR}_{s}/A^{NR}_{o}}-1)$ where $s$ correspond to the stoichiometric system (i.e. BaNbO$_{3}$ and Ca$_{6}$Al$_{7}$O$_{16}$, respectively) and $o$ are the corresponding off-stoichiometric system. As a comparison, we show the retarded relative surface correction at d$_s$=80\,nm in the right column.}
\end{table}

\begin{figure*}
 \includegraphics[width=0.9\textwidth,height=6.8cm]{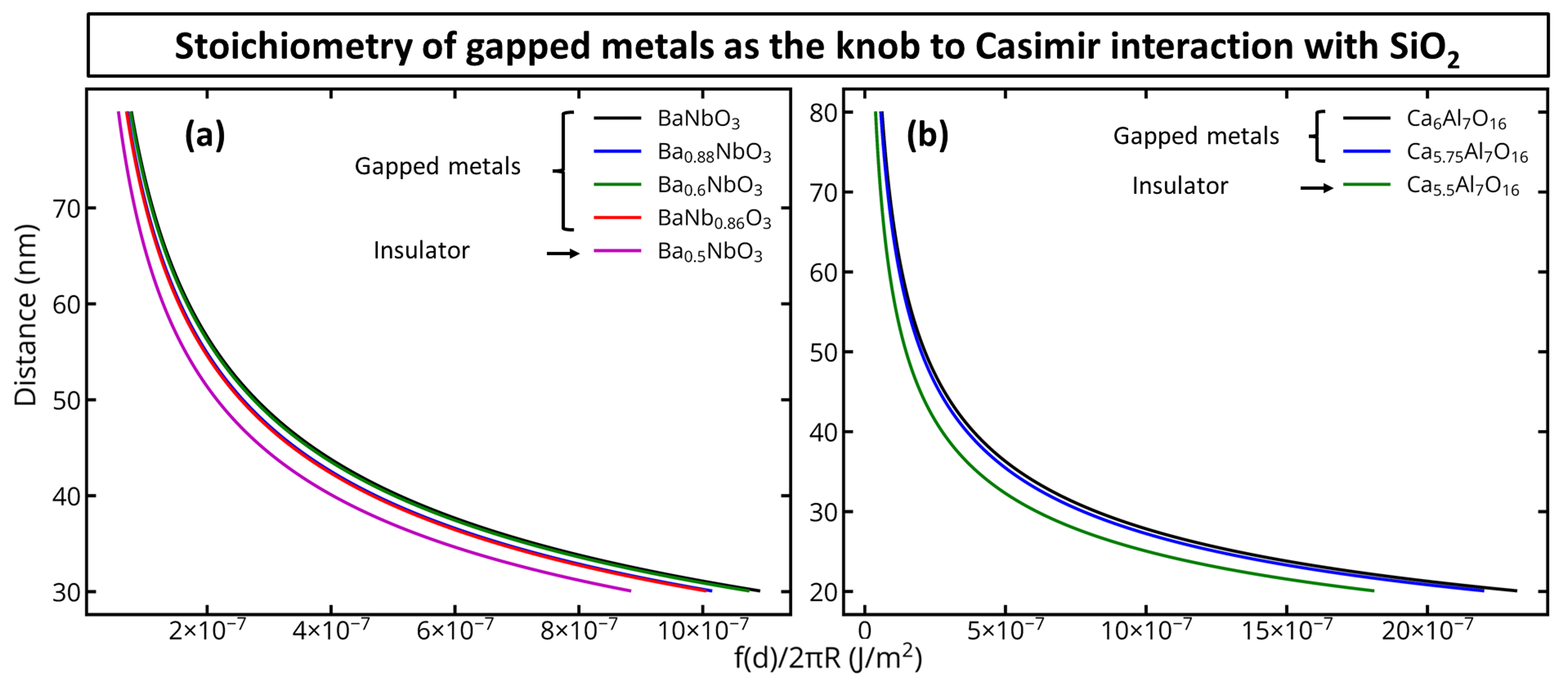}
 \vspace{-0.1in}
  \caption{\label{DistEnergy}   (Color online) Distance versus Casimir-Lifshitz force (obtained from proximity force theorem and the expression for the Casimir-Lifshitz free energy) using Eq.\,\ref{LifFreeEnergy}. In (a) the material combinations are Ba$_{1-x}$Nb$_{1-y}$O$_{3}$-air-SiO$_2$.  In (b) the material combinations are Ca$_{6-x}$Al$_{7}$O$_{16}$-air-SiO$_2$. }
\end{figure*}

 The relationship between the distance-dependent retarded Hamaker constant, $A^{\text{ret}}(d)$, and the corresponding free energy, $F(d, T)$, is expressed as $A^{\text{ret}}(d) = -12 \pi d^2 \times F(d, T)$. The retarded Hamaker constant between different silica-air-gapped metal systems are illustrated in Fig.\,\ref{RetardedHamaker1}. This figure illustrates how the interaction is tuned by using various material combinations of Ba$_{1-x}$Nb$_{1-y}$O$_{3}$-air-SiO$_2$.

This figure demonstrates how the interaction strength changes for different material combinations as the distances between particles vary. An attractive interaction can be determined by a positive retarded Hamaker constant, which decreases with increasing distance due to retardation. In the non-retarded (NR) limit of small distances ($d\rightarrow 0$) for each material combination, $A^{\text{NR}}$ takes a constant value, as listed in Tab.\,\ref{tbFittingParametersSubhojit}.
First, let us consider the limit of very small separations when retardation can be neglected. 
In this case, the Hamaker constant is truly a constant allowing us to calculate a relation between Hamaker constants and the deviations in surface separation ($\Delta d$),
\begin{equation}
{\frac{A_s (d_s)}{d_s^2}}={\frac{A_o (d_o)}{d_o^2}} \rightarrow  \Delta d=d_{s}-d_{o}\approx d_o(\sqrt{\frac{A_s}{A_o}}-1),
\label{DeltadFunction}
\end{equation}
and the percentage correction relative to the stoichiometric surface,
\begin{equation}
\delta_o=(d_s-d_o)\times100/d_o,
\label{deltao}
\end{equation}
\begin{equation}
\delta_o^{NR}=100\times(\sqrt{A^{NR}_{s}/A^{NR}_{o}}-1)
\label{deltao}
\end{equation}

The resulting percentage corrections in the non-retarded limit are given in Tab.\,\ref{tbFittingParametersSubhojit}.
The question is if these 1-7\% percentage distance corrections can give rise to corrections of the same order of magnitude as regular surface roughness.

To this aim, we plot, in a slightly unusual manner, the distance versus force for different Ba$_{1-x}$Nb$_{1-y}$O$_{3}$-air-SiO$_2$ in  Fig.\,(\ref{DistEnergy}\,a). Notably, the different distance curves correspond to different spontaneously formed stoichiometric or off-stoichiometric Ba$_{1-x}$Nb$_{1-y}$O$_{3}$ surface patches. Hence, for the same force different surface separations are predicted, corresponding to the tip (sphere) being above different surface patches.  To show the generality of the results we present a similar set of curves for Ca$_{6-x}$Al$_{7}$O$_{16}$-air-SiO$_2$ systems in Fig.\,(\ref{DistEnergy}\,b). We observe a clear trend for both systems when going from the most metallic to the most insulating surface material. 

\begin{figure}[h]
 \includegraphics[width=1\columnwidth]{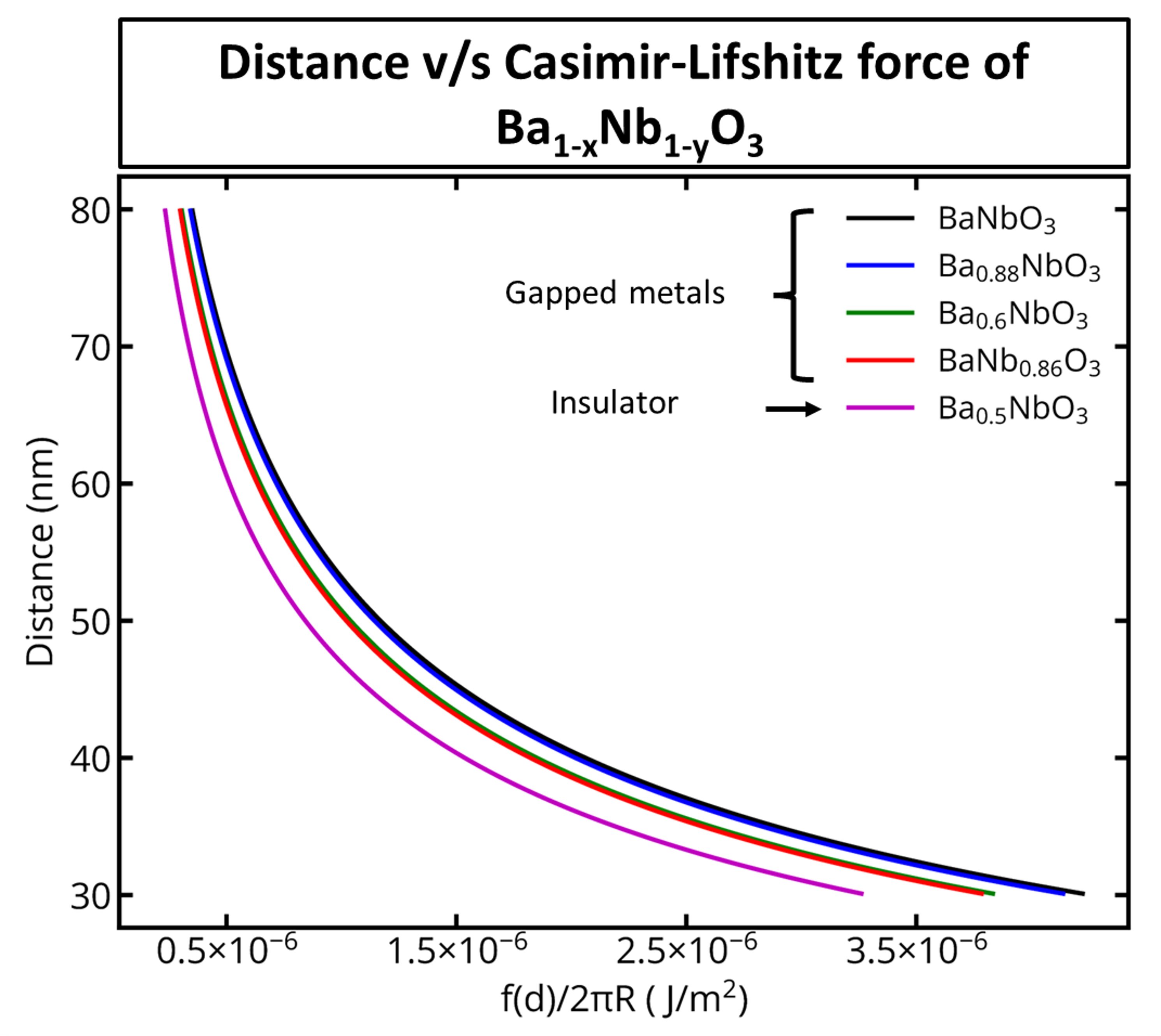}
  \vspace{-0.29in}
  \caption{\label{Gold1}   (Color online) Distance versus Casimir-Lifshitz force (obtained from proximity force theorem and the expression for the Casimir-Lifshitz free energy). The material combinations are Ba$_{1-x}$Nb$_{1-y}$O$_{3}$-air-Au.}
\end{figure}
\begin{figure}[h]
 \includegraphics[width=1\columnwidth]{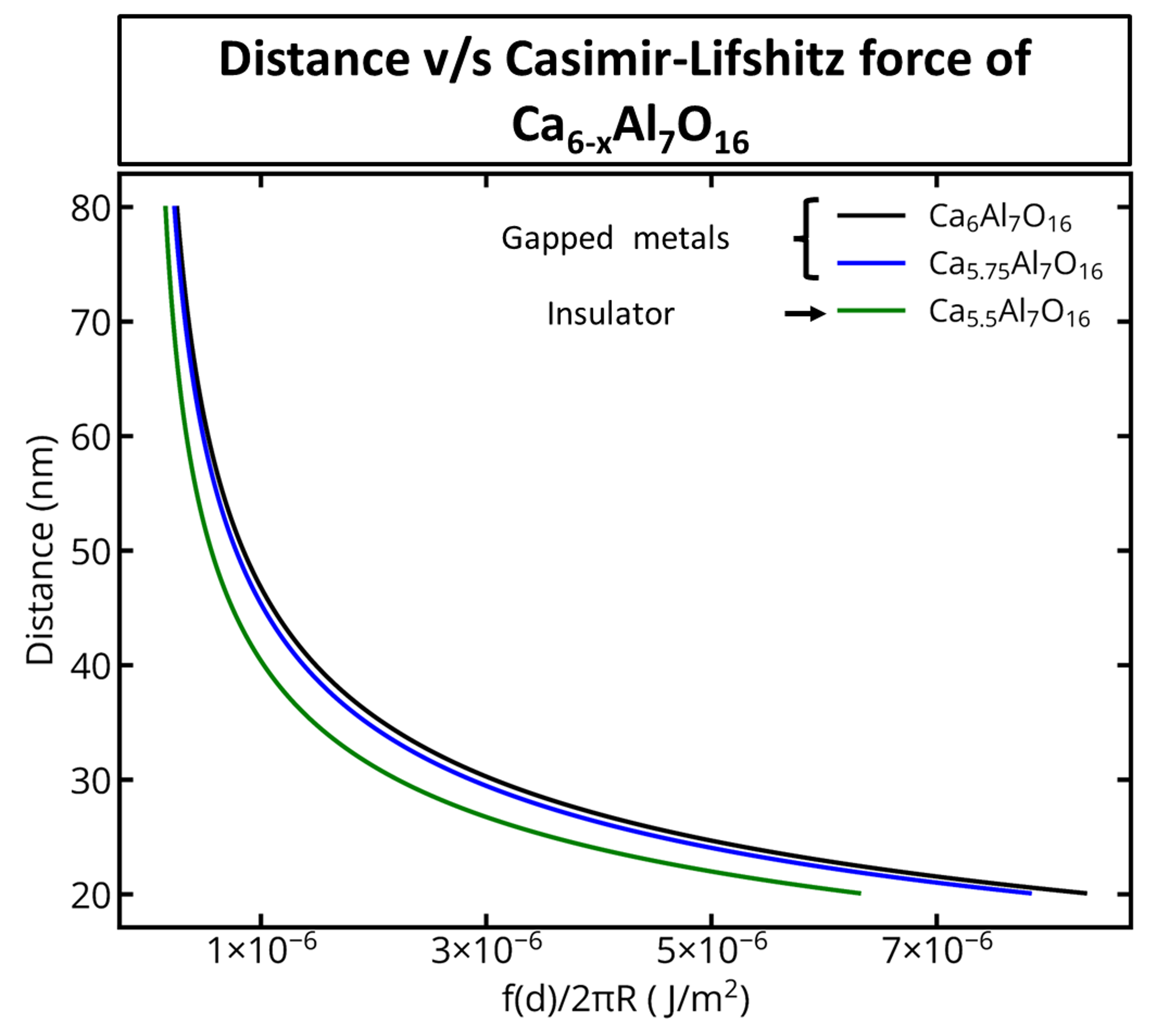}
  \vspace{-0.29in}
  \caption{\label{Gold2} (Color online) Distance versus Casimir-Lifshitz force (obtained from proximity force theorem and the expression for the Casimir-Lifshitz free energy). The material combinations are Ca$_{6-x}$Al$_{7}$O$_{16}$-air-Au.}
\end{figure}

\subsection{Gold systems} In parallel with calculations for SiO$_2$ spheres we also consider the case with a gold sphere. The dielectric function for gold was taken from Bostr\"om {\it et al.}\,\cite{BostromRizwanHarshanBrevikLiPerssonMalyi2023spontaneous}. The force curves for gold-air-gapped metals are shown in Fig.\,(\ref{Gold1}) and Fig.\,(\ref{Gold2}). Our proof-of-concept calculations considered a sphere made from either SiO$_2$ or gold.  However, we stress that for an insulating material, one can not apply any potential to perform electrostatic calibration. Hence, it is not possible to estimate contact potentials allowing for the application of a compensating potential during force measurements to minimize electrostatic contributions. As a result, the forces measured in an experimental setup with insulating materials can in principle be contaminated from uncompensated electrostatic contributions due to trapped charges\,\cite{babamahdi2019comparison}. This is a potential problem that limits any experimental verification of the theory predictions.

\subsection{Inverse design}
\begin{figure*}
 \includegraphics[width=0.9\textwidth,height=6.8cm]{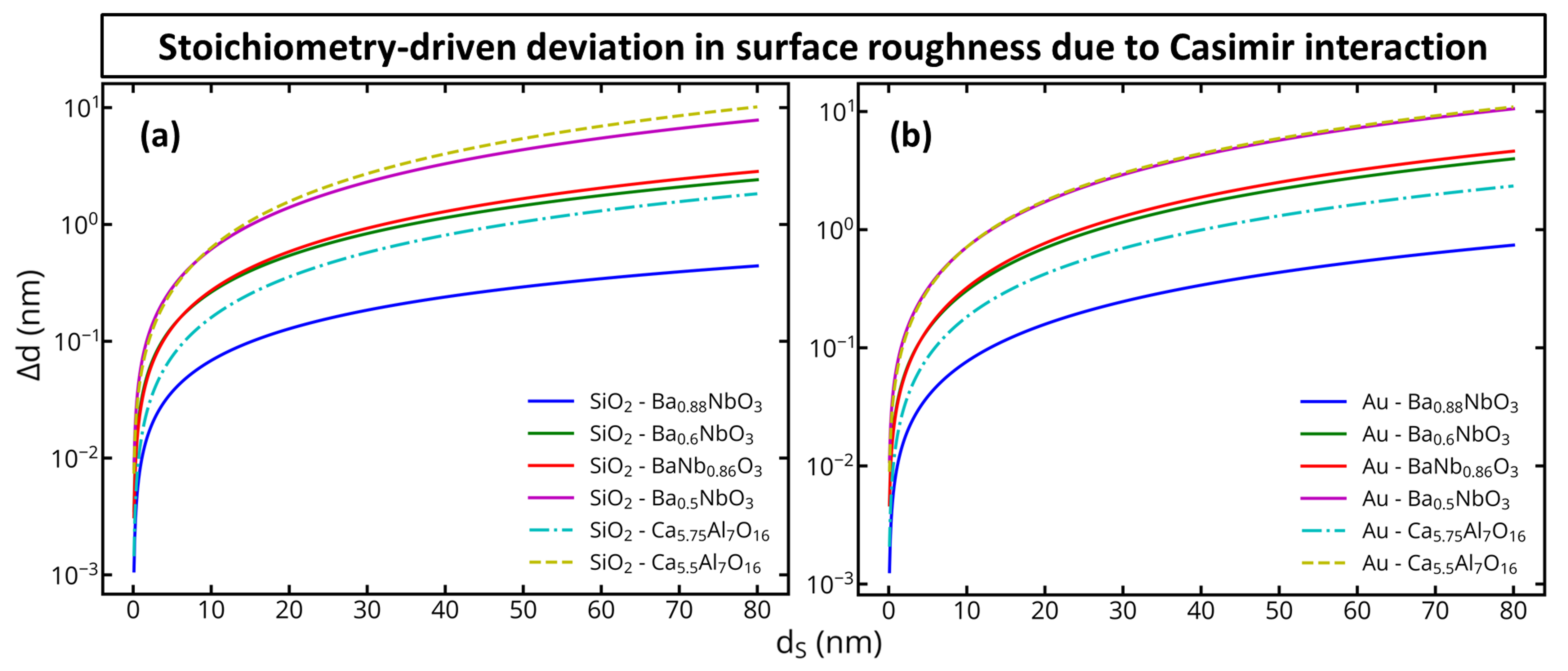}
 \vspace{-0.2in}
  \caption{\label{Deltadist}   (Color online) Here $\Delta\,d=d_s-d_o$ where $s$ correspond to the stoichiometric system (i.e. BaNbO$_{3}$ and Ca$_{6}$Al$_{7}$O$_{16}$, respectively) and $o$ are the corresponding off-stoichiometric system. In this figure, the fitted functions are used to evaluate the value of $\Delta\,d$ as a function of $d_s$. In figure (a) the sphere is made of silicon and in figure (b) it is made from gold. While the predicted forces are sensitive to the material of the sphere, we find that the distance variances have similar trends for silica and gold spheres.}
\end{figure*}

 Our Fig.\,(\ref{Deltadist}\,a) is one main result, and we obtain very similar results when the SiO$_2$ sphere is replaced with a gold sphere.  The results for gold shown in Fig.\,(\ref{Deltadist}\,b) demonstrate the generality of our inverse design approach as a gold-air-gapped metal system can be calibrated for potential electrostatic contributions.  The measured force is highly sensitive to both surface roughness and optical properties at separations less than 
100\,nm\,\cite{BroerPalasantzasPhysRevB.85.155410_2012}.  The ``inverse design'' used in the current work means that we study functions $d(f)$ rather than $f(d)$.  The fitting function for $d$ as a function of $x=f/[2 \pi R]$ can thus be written as $d = x^{a\ log(x)} \ x^b \ e^c$ where $a, b$ and $c$ are unitless fitted parameters given in Appendix.\,(\ref{appena}). These fitted functions are used in Fig.\,(\ref{Deltadist}\,a-b) to evaluate the value of distance variance, $\Delta\,d$ as a function of $d_s$, for different material combinations. This allows us to explore some surprisingly large corrections for the predicted surface separations. Notably, when comparing metallic patches with different stoichiometry, the corrections are around 1-4\%. However, even larger effects occur when we compare a stoichiometric metallic patch with a patch with an off-stoichiometric insulating state. We predict corrections up to 10-20\%. These effects are large enough to impact Casimir force measurements.

\section{Conclusions}
 In real-world applications, particularly for those involving heterogeneous surfaces, it is clear that a sphere attached to an AFM tip can result in significantly different measured forces when scanning over various surface regions. One should perhaps also mention that while gold systems also suffer from patch potential problems there are experimental methods used to work around that. Our key message is that differences in predicted sphere-surface separation are not only due to surface roughness or different electrostatic interactions but also partly due to heterogeneous surface patches.   In the case of gapped metal systems, these patches can spontaneously form during surface manufacturing.

\begin{acknowledgments}
MB, SP, and OIM's research contributions are part of the project No. 2022/47/P/ST3/01236 co-funded by the National Science Centre and the European Union's Horizon 2020 research and innovation programme under the Marie Sk{\l}odowska-Curie grant agreement No. 945339. 
Institutional and infrastructural
support for the ENSEMBLE3 Centre of Excellence was
provided through the ENSEMBLE3 project (MAB/2020/14)
delivered within the Foundation for Polish Science International
Research Agenda Programme and cofinanced by the
European Regional Development Fund and the Horizon 2020
Teaming for Excellence initiative (Grant Agreement No.
857543), as well as the Ministry of Education and Science
initiative "Support for Centres of Excellence in Poland under
Horizon 2020" (MEiN/2023/DIR/3797).
S.O. thanks the National Science Centre, Poland (grant no. UMO/2020/39/I/ST4/01446) and the "Excellence Initiative - Research University" (IDUB) Program, Action I.3.3 - "Establishment of the Institute for Advanced Studies (IAS)" for funding (grant no. UW/IDUB/2020/25). We gratefully acknowledge Poland's high-performance computing infrastructure PLGrid (HPC Centers: ACK Cyfronet AGH) for providing computer facilities and support within computational grant no. PLG/2023/016228 and for awarding this project access to the LUMI supercomputer, owned by the EuroHPC Joint Undertaking, hosted by CSC (Finland) and the LUMI consortium through grant no. PLL/2023/4/016319.  
\end{acknowledgments}

\appendix
\section{Fitting Functions and its analysis}
\label{appena}
Here, we discuss the fitting curve and parameters for distance vs. Casimir-Lifshitz force plot between different material combinations of  Ba$_{1-x}$Nb$_{1-y}$O$_3$ and Ca$_{6-x}$Al$_7$O$_{16}$ with Silica (SiO$_{2}$) and Gold (Au) sphere where the intermediate medium is air. We provide two Tab.\,\ref{tbFittingParametersSubhojit} and \ref{tbFittingParametersSubhojit1} for the above  combinations for the  fitted unitless parameters $a, b$ and $c$. These parameters are used to illustrate surface variance, $\Delta\,d$ as a function of $d_s$ in Fig.\,(\ref{Deltadist}a) and Fig.\,(\ref{Deltadist}b). These give us some information on insulating and metallic surface patches for different stoichoimetries. 

\begin{figure}[h]
 \includegraphics[width=1\columnwidth]{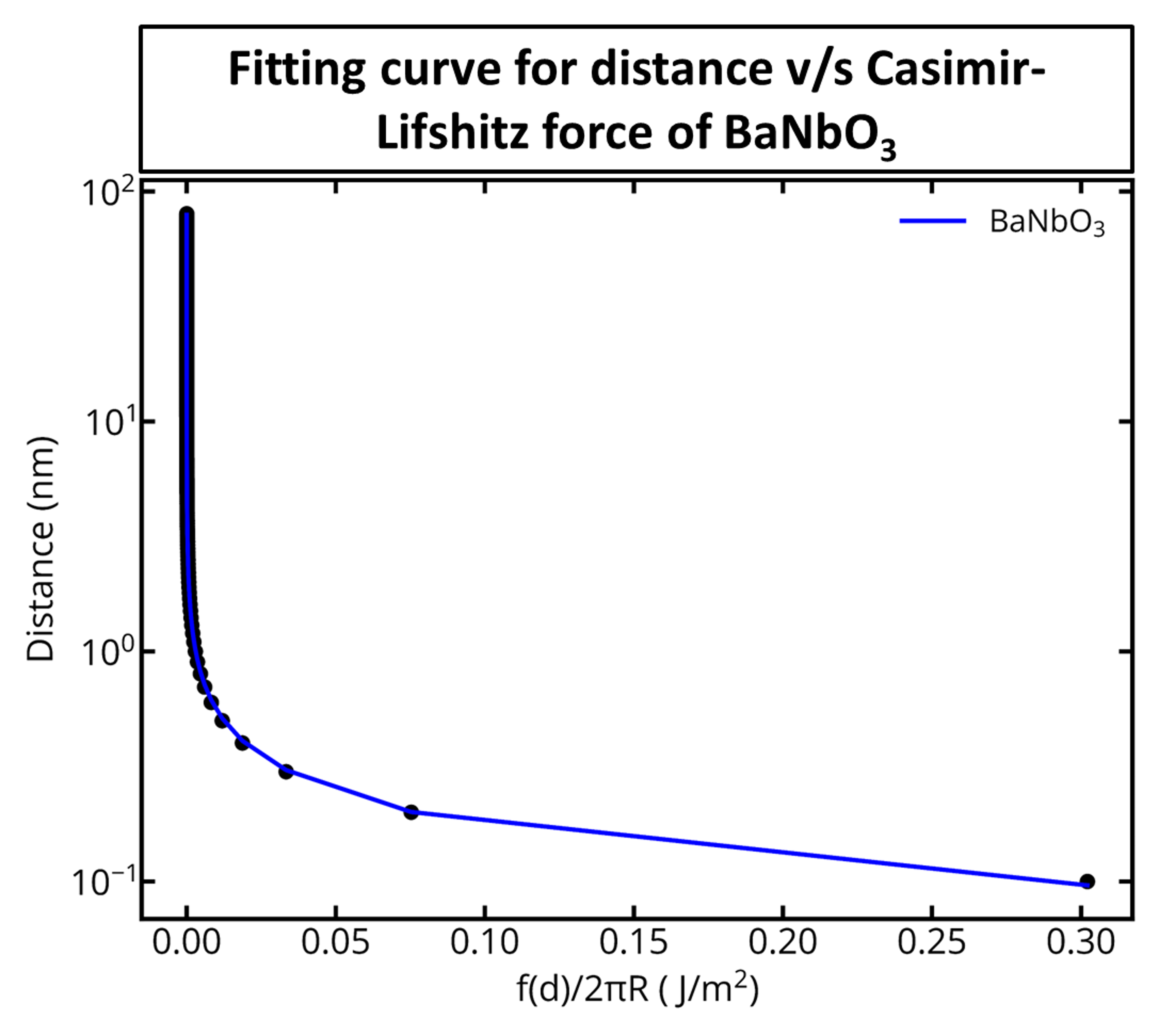}
  \vspace{-0.29in}
  \caption{\label{DielectricFunctions}   (Color online) Fitting curve for distance vs Casimir-Lifsitz force for BaNbO$_{3}$--air--SiO$_{2}$ system. }
\end{figure}
 \begin{table}[!h]
\centering
\begin{tabular}{c|c|c|c}
  \hline
   Material & a& b & c  \\
    \hline
 BaNbO$_{3}$ & -0.00624737 & -0.5521712 &-2.99345467\\
    \hline
     Ba$_{0.88}$NbO$_{3}$  &  -0.00636854 &-0.55473896& -3.03396482\\
    \hline
     Ba$_{0.6}$NbO$_{3}$  & -0.00626087& -0.55278433 &-3.00546641\\
    \hline
     BaNb$_{0.86}$O$_{3}$  &  -0.00638988 &-0.55445483 &-3.02918849\\
    \hline
     Ba$_{0.5}$NbO$_{3}$  &  -0.00689898 &-0.56260888 & -3.09362287\\
    \hline
     \hline
      \hline
      Ca$_{6}$Al$_{7}$O$_{16}$ &  -0.00598919& -0.55436462 &-3.18423815\\
    \hline
      Ca$_{5.75}$Al$_{7}$O$_{16}$ & -0.00607018 & -0.55530954 & -3.20351899 \\
    \hline
     Ca$_{5.5}$Al$_{7}$O$_{16}$  & -0.00703442  & -0.56918566 & -3.29720704 \\
    \hline
\end{tabular}
\caption{\label{tbFittingParametersSubhojit} The fitting function for silica--air--gapped metal surface  is $ y = x^{a\ log(x)} \ x^b \ e^c$ with $a, b$ and $c$ are unitless parameters given in the Table. }
\end{table}
\begin{table}[!h]
\centering
\begin{tabular}{c|c|c|c}
  \hline
   Material & a& b & c  \\
    \hline
 BaNbO$_{3}$ & -0.00496599 & -0.52243673 & -2.29460616\\
    \hline
     Ba$_{0.88}$NbO$_{3}$  &  -0.00502304 & -0.52352098 & -2.30748466\\
    \hline
     Ba$_{0.6}$NbO$_{3}$  &-0.00535562& -0.52810162 & -2.34418021\\
    \hline
     BaNb$_{0.86}$O$_{3}$  & -0.00542148 &-0.52840558 & -2.34266771\\
    \hline
     Ba$_{0.5}$NbO$_{3}$  & -0.00613957 & -0.53808893 & -2.41056251\\
    \hline
     \hline
      \hline
      Ca$_{6}$Al$_{7}$O$_{16}$ & -0.00507574 & -0.52958055 & -2.49814697\\
    \hline
      Ca$_{5.75}$Al$_{7}$O$_{16}$ & -0.00526731 & -0.53188494& -2.52399122\\
    \hline
     Ca$_{5.5}$Al$_{7}$O$_{16}$  & -0.00635692 & -0.54506319 &-2.61454029\\
    \hline
\end{tabular}
\caption{\label{tbFittingParametersSubhojit1} The fitting function for gold--air--gapped metal surface  is $ y = x^{a\ log(x)} \ x^b \ e^c$ with $a, b$ and $c$ are unitless parameters given in the Table. }
\end{table}

\section{Parameterised dielectric functions of the gapped metals}
\label{appenb}
In order to link optical calculations from DFT with force calculations, we parameterized the dielectric functions for the different gapped metals used in Tab.\,\ref{tab1} and\,\ref{tab2}.

\begin{table*}[h]
\centering
\caption{\label{tab1}Parametrization of the average dielectric function of continuous media, $\varepsilon(i\xi)$, at
imaginary frequencies for Ca$_{6-x}$Al$_{7}$O$_{16}$ as calculated with first-principles calculations and a damping coefficient ($\Gamma$) set to 0.2 \,eV. The $\omega_j$ modes are given in $\rm eV$. The largest difference between fitted and calculated $\varepsilon(i\xi)$ is 0.08\%.}

\begin{tabular}{ p{2.8cm} | p{2.8cm} | p{2.8cm} | p{2.8cm} }
\hline
\hline
{modes ($\omega_j$)}  & \multicolumn{3}{c}{ Coefficients ($C_j$) for different Ca$_{6-x}$Al$_{7}$O$_{16}$ compounds}  \\
\hline
 & Ca$_{6}$Al$_{7}$O$_{16}$ & Ca$_{5.75}$Al$_{7}$O$_{16}$ & Ca$_{5.5}$Al$_{7}$O$_{16}$ \\
\hline
0.0206 & 58.9601 & 0.6494 & 0.0001 \\
\hline
0.0347 & 91.1774 & 1.797 & 0.0003 \\
\hline
0.0587 & 57.4068 & 5.2997 & 0.001 \\
\hline
0.1013 & 16.4729 & 15.4951 & 0.0221 \\
\hline
0.1996 & 73.0451 & 29.8463 & 0.3283 \\
\hline
0.3938 & 0.3949 & 2.1519 & 0.7511 \\
\hline
0.9556 & 0.0706 & 0.2519 & 0.4345 \\
\hline
2.2773 & 0.0987 & 0.0392 & 0.0 \\
\hline
6.4732 & 0.4594 & 0.2384 & 0.2279 \\
\hline
10.2048 & 0.7938 & 0.8189 & 0.0 \\
\hline
18.2421 & 0.3705 & 0.474 & 0.0486 \\
\hline
30.9018 & 0.1655 & 0.2388 & 0.0 \\
\hline
54.455 & 0.0059 & 0.0283 & 0.001 \\
\hline
\hline
\end{tabular}
\end{table*}

\begin{table*}[h]
\centering
\caption{\label{tab2}Parametrization of the average dielectric function of continuous media, $\varepsilon(i\xi)$, at
imaginary frequencies for Ba$_{1-x}$Nb$_{1-y}$O$_{3}$ as calculated with first-principles calculations and a damping coefficient ($\Gamma$) set to 0.2 \,eV. The $\omega_j$ modes are given in $\rm eV$. The largest difference between fitted and calculated $\varepsilon(i\xi)$ is 0.1\%.}

\begin{tabular}{ p{2.1cm} | p{2.1cm} | p{2.1cm} | p{2.1cm} | p{2.1cm} | p{2.1cm}}
\hline
\hline
{modes ($\omega_j$)}  & \multicolumn{5}{c}{Coefficients ($C_j$) for different Ba$_{1-x}$Nb$_{1-y}$O$_{3}$ compounds}  \\
\hline
 & BaNbO$_{3}$ & BaNb$_{0.88}$O$_{3}$ & Ba$_{0.6}$NbO$_{3}$ & BaNb$_{0.86}$O$_{3}$ & Ba$_{0.5}$NbO$_{3}$\\
\hline
0.0215 & 9.4105 & 69.1791 & 10.1195 & 0.6139 & 0.0 \\
\hline
0.0438 & 17.6756 & 94.1209 & 15.8813 & 4.5368 & 0.0 \\
\hline
0.0872 & 19.2732 & 66.833 & 17.5738 & 7.3998 & 0.0 \\
\hline
0.1967 & 451.2032 & 351.9936 & 83.8965 & 80.6752 & 0.0 \\
\hline
0.2227 & 53.9568 & 37.1564 & 35.407 & 0.0 & 0.0 \\
\hline
0.6448 & 0.2725 & 1.3619 & 1.6775 & 6.027 & 0.0 \\
\hline
2.5296 & 0.1729 & 0.1616 & 0.3952 & 0.6325 & 0.2161 \\
\hline
5.0772 & 1.2805 & 1.5587 & 2.4751 & 1.2681 & 2.4029 \\
\hline
8.4405 & 1.5742 & 1.3843 & 0.8565 & 1.7721 & 1.0033 \\
\hline
16.0637 & 0.7289 & 0.7517 & 0.8088 & 0.5311 & 0.717 \\
\hline
27.1476 & 0.2407 & 0.2107 & 0.1615 & 0.3452 & 0.1712 \\
\hline
47.1511 & 0.0632 & 0.0695 & 0.0824 & 0.0074 & 0.0782 \\
\hline
79.6918 & 0.0029 & 0.0 & 0.0 & 0.0114 & 0.0 \\
\hline
\hline
\end{tabular}
\end{table*}
\newpage



\end{document}